\documentclass[letterpaper, 10pt, conference]{IEEEtran}
\usepackage{latexsym}
\usepackage{amsmath}
\usepackage{amssymb}
\usepackage{amsfonts}
\usepackage{graphicx}
\usepackage{pstricks}
\usepackage{pst-node,pst-plot}
\usepackage{setspace}
\usepackage{hyperref}
\newcommand{\m}[1]{\ensuremath{\mathbf{#1}^n}}
\newcommand{\neigh}[1]{\ensuremath{\mathbf{Y}_{\sim #1}^{(l)}}}

\title{Tradeoff between decoding complexity and rate for codes on graphs}
\author{Pulkit Grover\\ Wireless Foundations, EECS Department\\ University of California at Berkeley\\ Email: pulkit\; @\; eecs.berkeley.edu}
\begin{document}\maketitle
\begin{abstract}
%THIS PAPER IS ELIGIBLE FOR THE STUDENT PAPER AWARD\\
We consider transmission over a general memoryless channel, with bounded decoding complexity per bit under message passing decoding. We show that the achievable rate is bounded below capacity if there is a finite success in the decoding in a specified number of operations per bit at the decoder for some codes on graphs. These codes include LDPC and LDGM codes. Good performance with low decoding complexity suggests strong local structures in the graphs of these codes, which are detrimental to the code rate asymptotically. The proof method leads to an interesting necessary condition on the code structures which could achieve capacity with bounded decoding complexity. We also show that if a code sequence achieves a rate $\epsilon$ close to the channel capacity, the decoding complexity scales at least as $O\left(\log\left(\frac{1}{\epsilon}\right)\right)$.

%: the marginal uncertainty in knowledge of any codeword bit should \textit{not} decrease for the codes to achieve capacity with bounded decoding complexity.  
\end{abstract}
\section{Introduction}
In \cite{khandekarthesis} and \cite{mceliece}, Khandekar and McEliece suggested studying the problem the problem of \textit{per bit} decoding complexity for capacity achieving codes. The authors conjectured on some lower bounds for the asymptotic decoding complexity for a sequence of codes achieving a rate which is $\epsilon$ close to the channel capacity. For general memoryless channels, with message passing decoding, the authors conjectured a lower bound of $O(\frac{1}{\epsilon}\log\left(\frac{1}{\epsilon}\right))$ operations per bit at the decoder. There is one exception: for the BEC, codes were already known \cite{amincapacity} which achieve capacity with $O\left(\log\frac{1}{\epsilon}\right)$ operations per bit at the decoder (see \cite{sason}). The BEC is an exception because along each edge in the graph, a message needs to be passed only once.

Soon after,  in \cite{UrbankeSason}\cite{anastapopulis} the authors constructed codes which achieve capacity with bounded decoding complexity for the BEC. In \cite{PfisterSason}, the authors construct systematic Accumulate-Repeat-Accumulate codes which achieve capacity with bounded decoding complexity. Interestingly, all the known codes which achieve capacity with bounded decoding complexity use accumulation.

It is a natural question whether there exist codes which achieve capacity with bounded decoding complexity under message passing decoding for general memoryless channels. It is a hard question, too, for we yet do not know of any codes which achieve capacity under message passing decoding for general memoryless channels.

In this paper, using an insight developed in \cite{etesami}, we show that a lower bound of $O(\log(\frac{1}{\epsilon}))$ holds for a large class of codes, including LDPC codes and LDGM codes, and any sequence of systematic codes. 

This paper is organized as follows. In Section \ref{sec:notation} we introduce the notation and some definitions used in this article. In Section \ref{sec:ourresults} we put forward our lower bound on the decoding complexity. We also find some bounds on the rate of codes given their decoding performance. In Section \ref{sec:exist} we observe a necessary condition for codes to achieve capacity with bounded decoding complexity: for infinite block-length, there should be no decrease in probability of error in finite number of operations per bit at the decoder. Section \ref{sec:conclusions} concludes the paper.

%%%%%%%%%%%%%%%%%%%%%%%%%%%%%%
\section{Notation and Definitions}
\label{sec:notation}
The block length of a code is denoted by $n$. A vector comprising of first $i$ elements of a sequence of random variables is denoted in bold letters with superscript
$i$. For example, the channel input vector is denoted by
$\mathbf{X}^n:=\{X_1,X_2,\ldots,X_n\}$. Similarly, the 
 channel output vector is denoted by $\mathbf{Y}^n$. $k^{th}$ element of any vector $\mathbf{X}^n$ is denoted by $X_k$. It would also be convenient to define $\mathbf{Y}_{\sim i}^{(l)}$ as the outputs of the variable nodes in the neighborhood of node $i$ till the $l^{th}$ iteration (excluding node $i$). $C$ is used to denote the capacity of the channel under consideration. We assume that the channel is binary input, symmetric, and memoryless, but is otherwise arbitrary. 
 
 Let $\epsilon$ be the gap from capacity, $R=C-\epsilon$. Denote the `success in decoding' by a parameter $\tau$. $\tau$ may correspond to decrease in probability of error, or a similar parameter characterizing some success in decoding\footnote{Later, we use the decrease in entropy of an output symbol from $H(Y_i)$ as the parameter. See \eqref{eq:givenbound}.}.

We denote the number of decoding operations required per information bit for decoding a sequence of codes by $\chi_D^{(\tau)}(\epsilon)$, where $\epsilon$ is the difference between the code rate $R$ and the channel capacity $C$.

All the results in this paper hold for general memoryless channel holds with one exception, the BEC.
%%%%%%%%%%%
%%%%%%%%%%
\section{An upper bound on the rate and a lower bound on the decoding complexity}
\label{sec:ourresults}
Suppose we have a sequence of LDPC codes of fixed degree distributions $(\lambda, \rho)$. In asymptotic (large block-length) analysis, the probability density of the messages passed \cite{urbankecapacity} converges in distribution to $\delta_\infty$, a point mass at infinity, as the number of iterations $l$ converges to infinity. A message to $i^{th}$ variable node is a log likelihood ratio $\frac{\Pr(X_i=0|\neigh{i})}{\Pr(X_i=1|\neigh{i})}$. Therefore, the random variable $\Pr(X_i=0|\neigh{i})\overset{d}{\rightarrow} 1$, and $H(X_i|\neigh{i})\overset{d}{\rightarrow}0$ as $l\rightarrow\infty$. The randomness is over the channel realization, the choice of the code in the ensemble, and the choice of the node.

Given some success in decoding by the $l^{th}$ iteration, with high probability the conditional entropy of the $i^{th}$ output bit $Y_i$ can then be bounded as follows
\begin{equation}
\label{eq:givenbound}
H(Y_i|\neigh{i})<H(Y_i)-\tau_l
\end{equation}
This was first observed in \cite{etesami} in the context of LT codes. 

For some finite decoding success in constant number of computations, we prove that the rate is bounded below channel capacity, by explicitly finding the bound. We then arrive at a lower bound on the decoding complexity for a code being used at rate $R=C-\epsilon$. For derivation of these bounds, we first need a bound on $H(\m{Y})$.

\subsection{Bound on $H(\m{Y})$}
Consider the decoding up to $l$ iterations. Let $\epsilon_l$ denote the probability of error after $l$ iterations. 

After $l$ iterations, the conditional entropy $H(Y_i|\mathbf{Y}_{\sim i}^{(l)})$ can be bounded above by
\begin{equation}
H(Y_i|\mathbf{Y}_{\sim i}^{(l)})<H(Y_i)-\tau_l
\end{equation}
%Problem of notation. Y^l and Y^n.

Using this bound, we wish to upper bound the entropy of vector $\mathbf{Y}^n$. Using the chain rule
\begin{equation}
\label{eq:chain}
H(\mathbf{Y}^n)= \sum_{i=1}^n H(Y_i|\mathbf{Y}^{i-1})
\end{equation}
Naturally enough, we would like to use the fact that conditioning reduces entropy to upper bound the individual conditional entropies in the sum.  To that end, we want to remove some elements of the vector $\mathbf{Y}^{i-1}$ in order to arrive at $\mathbf{Y}_{\sim i}^{(l)}$. But notice that this would not always be possible. In fact, it is not difficult to show that in general, no particular permutation of the vector $\mathbf{Y}$ can ensure that $\mathbf{Y}_{\sim i}^{(l)} \subset \mathbf{Y}^{i-1}$ for a constant fraction of $Y_i$'s.

In \cite{etesami}, this problem is dealt with by showing\footnote{Strictly speaking, in \cite{etesami}, the bound is on mutual information. But that can easily be translated to a bound on conditional entropies that we are dealing with.} that the set $\mathbf{Y}_{\sim i}^{(l)} \backslash \mathbf{Y}^{i-1}$ can be considered to be erased for the analysis, and they try to find another constant $\tau'$ which bounds the entropy $H(Y_i|\mathbf{Y}_{\sim i}^{(l)}\cap \mathbf{Y}^{i-1})$. In the following, we show a neater way to bound $H(\mathbf{Y})$ which directly leads to a clean upper bound on the rate. Moreover, this method works for LDPC codes as well as some other classes of codes, which is not the case for the the method in \cite{etesami}.

Consider all the $n!$ permutations of $\mathbf{Y}^n$. Consider the set $\mathbf{Y}_{\sim i}^{(l)}$, constituted by the output values of the bits corresponding to the nodes in the neighborhood after $l$ iterations. For each $i$, this set has a constant number of elements equalling $k_l$ (say). It is important to note that for a fixed sequence of codes, $k_l$ is independent of $n$. Because this set is constant in size, we try to find the fraction of permutations for which this set occurs before the particular $Y_i$ we are considering. The problem is equivalent to finding the number of permutations of a set in which a particular element of a subset occurs last amongst all the elements of the subset. This would be true for exactly $\frac{1}{k_l}$ fraction of total permutations, since it is the same for each element of the subset.

The size of the neighborhood set $k_l\approx \sum_i^l(\alpha-1)^i(\beta-1)^i$, where $\alpha$ ($\beta$) is the average variable (check) node degree. Also, $\frac{\alpha}{\beta}=1-R$, where $R$ is the code rate.

We now derive the bound. In the following the bracketed $(j)$ represents the $j^{th}$ permutation ordering of the vector $\mathbf{Y}^n$.

\begin{eqnarray*}
\frac{1}{n}H(\mathbf{Y}^{n})&=&\frac{1}{n\times n!}\sum_{j=1}^{n!} H(\mathbf{Y^n(j)})
\\&=&\frac{1}{n\times n!}\sum_{j=1}^{n!}\sum_{i=1}^nH(Y_i(j)|\mathbf{Y}^{i-1}(j))
\end{eqnarray*}
Exchanging the two summations
\begin{eqnarray*}
\frac{1}{n}H(\mathbf{Y}^{n})=\frac{1}{n}\sum_{i=1}^{n}\frac{1}{n!}\sum_{j=1}^{n!}H(Y_i(j)|\mathbf{Y}^{i-1}(j))
\end{eqnarray*}
As argued in the previous paragraph, the terms in the summation can be bounded above by $H(Y_i)-\tau$ for $\frac{1}{k_l}$ fraction of these permutations. Therefore,
\begin{eqnarray*}
\frac{1}{n}H(\mathbf{Y}^{n})=\frac{1}{n}\sum_{i=1}^{n} \big( \frac{1}{k_l}(H(Y_i)-\tau_l) + (1-\frac{1}{k_l})H(Y_i)\big)
\end{eqnarray*}
Since $Y_i$'s are identically distributed, we get the bound
\begin{equation}
\label{eq:permutebound}
\frac{1}{n}H(\mathbf{Y}^{n})\leq H(Y_i)-\frac{\tau_l}{k_l}
\end{equation}
Per se, this bound is very loose, since the fraction $\frac{1}{k_l}$ converges to 0 polynomially in the average degree. 

%Notice that the RHS of \eqref{eq:permutebound} depends on $R$, because of the dependance of $\beta$ on $R$. But this dependance can be dealt with by finding the supremum over $R$ for which this inequality is satisfied. A clearer (though not as tight) way would be to lower bound $1-R$ by $1-C$, since $R<C$.

In comparison with \cite{etesami}, this bound is explicit on its dependance on the degree distribution. In the following section, we show how this bound leads to a bound on the achievable rate for LDPC codes. 
%It is important to note that the bound in \cite{etesami} is dependent on the degree distribution.

%%%%%%%%%%%%%%%%%%%%%%%%%%%%%%%%
\subsection{Bound on the achievable rate for given performance after a fixed number of iterations}
Using Fano's inequality,
\begin{equation}
\label{eq:fano}
H(P_e)\geq \frac{1}{n}H(\mathbf{X}^n|\mathbf{Y}^n)
\end{equation}
Using standard information-theoretic equalities
\begin{equation}
\label{eq:galeq1}
\frac{1}{n}H(\mathbf {X}^n|\mathbf {Y}^n) =\frac{1}{n}H(\mathbf {X}^n)- \frac{1}{n}H(\mathbf {Y}^n)+\frac{1}{n}H(\mathbf {Y}^n|\mathbf {X}^n)
\end{equation}
For any code, $\mathbf{X}^n$ is the encoded data which is in 1-1 mapping
with the information symbols. The information
symbols are uniformly distributed over the $2^{nR}$ values. Thus,
$\mathbf{X}^n$ takes any value from the $2^{nR}$ codewords with
uniform probability distribution. Therefore 
\begin{equation}
\label{eq:x}
H(\mathbf{X}^n)=nR
\end{equation}
Using the fact that the channel is memoryless
\begin{equation}
\frac{1}{n}H(\mathbf{Y}^n|\mathbf{X}^n)=H(Y_i|X_i)=H(Y_i)-C
\end{equation}
Notice that in \eqref{eq:galeq1}, it now suffices to upper bound the
entropy $H(\mathbf{Y}^n)$, which is what we found in \eqref{eq:permutebound}. Along with \eqref{eq:fano} we obtain
\begin{equation}
\label{eq:usingfano}
H(P_e)\geq R-\big( H(Y_i)-\frac{\tau_l}{k_l}\big)+H(Y_i)-C=R+\frac{\tau_l}{k_l}-C
\end{equation}
Therefore, for $P_e\rightarrow 0$, we get the following bound on the achievable rate
\begin{equation}
\label{eq:boundonrate}
R\leq C-\frac{\tau_l}{k_l}
\end{equation}
Also, notice that \eqref{eq:usingfano} gives us a lower bound on the probability of error for the case when this bound is violated.
\subsubsection{Tightening the bound with more knowledge of the code performance}
Since the bound is valid for all iterations $l$, we get the following tighter bound
\begin{equation}
\frac{1}{n}H(\mathbf{Y}^{n})\leq \underset{l}{\min}\left\{H(Y_i)-\frac{\tau_l}{k_l}\right\}
\end{equation}
where the minimum is taken over all the values of $l$ for which the decoding performance is known.

%The bound can further be tightened for performance known for more than one iterations. But we skip the tightening since that is not central to the discussion.
%%%%%%%%%%%%%%%%%%%%%%%
\subsection{Bound for given decoding success in fixed number of computations}
Denote the neighborhood size by $k^{(c)}$ (which depends on $l$ and $\alpha$), and the success in decoding by $\tau^{(c)}$. Then it is easy to see that the same derivation still works, and we get the following bound on entropy of $\m{Y}$
\begin{equation}
\label{eq:forc}
\frac{1}{n}H(\m{Y})\leq H(Y_i)-\frac{\tau^{(c)}}{k^{(c)}}
\end{equation}
The total number of operations at the decoder is $(\alpha \times l+\beta\times l)n$. Therefore, the number of operations per information bit,
\begin{equation}
c = \frac{(\alpha +\beta)l}{R}
\end{equation}
%The denominator in above equation does not converge to infinity, as the average degree is increased (in fact, it converges to 1). 
Define $k_{\mbox{max}}$ as the maximum value of $k^{(c)}$ for fixed c.
\begin{equation}
k_{\mbox{max}}=\underset{l\geq 1}{\sup}\left\{\sum_i^l (\alpha-1)^i(\beta-1)^i\right\}
\end{equation}
%where $\alpha=\frac{c}{l\geq 1}$, and $\frac{\alpha}{\beta}=1-R$. Because of the dependance of $\beta$ on $R$, the inequality~\eqref{eq:forc} is again dependent on $R$ itself. Using $1-R>1-C$, 
Then, 
\begin{eqnarray*}
k_{\mbox{max}}&\leq& \underset{l\geq 1}{\sup}\left\{\sum_{i=1}^l (\alpha+\beta-1)^{2i}\right\}
\\& \leq  & \underset{l\geq 1}{\sup}\left\{ (\alpha + \beta-1)^{2l+2}\right\}
\\ & \leq & \underset{l\geq 1}{\sup}\left\{ (\alpha + \beta)^{2l+2}\right\}
\\ & = & \underset{l\geq 1}{\sup}\left\{ \left(\frac{Rc}{l}\right)^{2l+2}\right\}
\\ &\leq & \underset{l\geq 1}{\sup}\left\{ \left(\frac{c}{l}\right)^{2l+2}\right\}
\end{eqnarray*}
since $R\leq 1$. Define 
\begin{equation}
k_{bd}^{(c)}: = \underset{l\geq 1}{\sup} \left\{\left( \frac{c}{l}\right)^{2l+2}\right\} 
\end{equation}
Then $k_{bd}^{(c)}<\infty$ for each finite $c$ because as $l\rightarrow 1$, $k_{bd}^{(c)}\rightarrow (c)^4$, and as $l\rightarrow\infty$, $k_{bd}^{(c)}\rightarrow 0$.

Thus a bound on the achievable rate is
\begin{equation}
\label{eq:bd}
R\leq C-\frac{\tau^{(c)}}{k_{bd}^{(c)}}
\end{equation}
Since $k_{bd}^{(c)}<\infty $, $R$ is bounded below capacity. 
\subsection{A lower bound on decoding complexity as a function of gap from capacity}
Substituting $R=C-\epsilon$ in \eqref{eq:bd},
\begin{equation}
k_{bd}^{(c)} \geq \frac{\tau^{(c)}}{\epsilon}
\end{equation}  
Numerically, we observed that $k_{bd}^{(c)} < e^{c}$ for $c$ large enough. Therefore, for fixed $\tau$, the per bit decoding complexity
\begin{equation}
\chi_D^{(\tau)}\geq O\left(\log\left(\frac{1}{\epsilon}\right)\right)
\end{equation}
Here $\tau$ is the parameter which reflects the success in decoding in \eqref{eq:givenbound}.

\subsection{Generalization to other codes}
Notice that we did not use any structure of LDPC codes for deriving the bound above. All we require is some finite success in decoding the output bits in order to decode the information bits. The bounds above, therefore, generalize to LDGM codes.

If we assume plain message passing decoding, then for any systematic code, there has to be a finite success in the decoding of the information bits with a finite neighborhood size. This implies a finite success in the decoding of output bits! Therefore, no systematic code can achieve capacity with bounded decoding complexity under plain message passing decoding for general memoryless channels. 

\section{A necessary condition for codes which achieve capacity with bounded decoding complexity}
\label{sec:exist}
Note that the results in Section \ref{sec:ourresults} are not valid for the BEC. It is shown in \cite{UrbankeSason}\cite{anastapopulis} that there exist codes which achieve capacity with bounded decoding complexity for the BEC. Considering the above bound, it is intriguing that such codes exist. After all, the above bound only depends on \textit{some} knowledge of marginal of output bits. As long as the probability of the output bits being 1 is bounded away from 0.5, the bound holds. 

The codes in \cite{UrbankeSason} are non-systematic IRA codes. IRA codes have an LDGM inner code, and a convolution code as the outer code. The convolution structure means that for any particular bit, for any finite number of iterations $l$, the marginal probability $\Pr(Y_i=1|\neigh{i})=0.5$. The marginal uncertainty of the output bits does not decrease with increase in $l$! This also hints why systematic IRA codes, for which some of the output bits are decoded in finite time, fail to achieve capacity with bounded decoding complexity. The accumulation operation results in differential encoding of the inner code. The marginal uncertainty of the output bits does not decrease. But the joint uncertainty decreases substantially, giving us an estimate of the inner code bits, and therefore the information bits.

The arguments above are not tight for the BEC, because each edge in the graph needs to be used only once for the BEC. However, they hold for general memoryless channels. The bounds in Section~\ref{sec:ourresults} therefore imply that an accumulation-like operation is necessary for any sequence of codes to achieve capacity with bounded decoding complexity. Accumulation is really an example of a convolution code. Therefore, an outer convolution code would mean that the bounds in Section~\ref{sec:ourresults} do not apply.

\section{Discussions and Conclusions}
\label{sec:conclusions}
We showed that under message passing decoding, the achievable rate of a sparse graph code sequence is bounded below capacity if there is a finite success with bounded complexity decoding. The bound in Section \ref{sec:ourresults} is actually because a bounded decoding complexity also bounds the neighborhood size. Some success in decoding with a bounded neighborhood size suggests strong local structures in the code, which is detrimental to its performance.

The bounds imply that an accumulation-type outer code would be necessary in order to achieve capacity with bounded decoding complexity for general memoryless channels.
% The codes which achieve

%The results need to be translated in terms of probability of error, instead of the gap $\tau_{\alpha l}$. This should not be difficult for particular channels, though a general result may use bounding techniques.

%The bound on the rate can certainly be tightened for specific cases. In particular, using methods in \cite{etesami} the bound found (for LT codes) should be stronger, since they do not neglect the case when $\mathbf{Y}_{\sim i}^{(l)}\nsubseteq \mathbf{Y}^{i-1}$. Though we have not investigated whether these bounds are stronger.

%The method of derivation is fairly general. Hence it appears that it can be useful in other areas as well. For example, it's worthwhile investigating whether systems for which Belief Propagation converges very fast are sub-optimal in some sense. In coding theory, it would be interesting to explore LP decoding \cite{wainwright} to see if the bounds hold there.

\bibliographystyle{unsrt}
\bibliography{281project}

\begin{thebibliography}{1}

\bibitem{khandekarthesis}
A~Khandekar.
\newblock {\em Graph-based codes and iterative decoding}.
\newblock PhD thesis, California Institute of technology, Pasadena, CA, 2002.

\bibitem{mceliece}
A~Khandekar and RJ~McEliece.
\newblock {On the complexity of reliable communication on the erasure channel}.
\newblock In {\em IEEE International Symposium on Information Theory}, 2001.

\bibitem{amincapacity}
Amin Shokrollahi.
\newblock Capacity achieving sequences.
\newblock In {\em Codes, Systems and Graphical Models}, number 123 in IMA
  Volumes in Mathematics and its Applications, pages 153--166, 2000.

\bibitem{sason}
Igal Sason and RL~Urbanke.
\newblock Parity-check density versus performance of binary linear block codes
  over memoryless symmetric channels.
\newblock {\em IEEE Transactions on Information Theory}, 49(7):1611--1635, July
  2003.

\bibitem{UrbankeSason}
H~Pfister, I~Sason, and RL~Urbanke.
\newblock {Capacity-Achieving Ensembles for the Binary Erasure Channel With
  Bounded Complexity}.
\newblock In {\em IEEE International Symposium on Information Theory}, page
  209, 2004.

\bibitem{anastapopulis}
CH~Hsu and A~Anastasopoulos.
\newblock Capacity-achieving codes with bounded graphical complexity on noisy
  channels.
\newblock In {\em Proc. 43th Allerton Conf. on Communication, Control, and
  Computing}, Monticello, Illinois, USA, October 2005.

\bibitem{PfisterSason}
H.~D. Pfister and I.~Sason.
\newblock Accumulate-repeat-accumulate codes: Systematic codes achieving the
  binary erasure channel capacity with bounded complexity.
\newblock {\em {\textit{submitted to}} IEEE Transactions on Information
  Theory}, 2005.

\bibitem{etesami}
O~Etesami, M~Molkaraie, and MA~Shokrollahi.
\newblock Raptor codes on symmetric channels.
\newblock {\em {\textit{submitted to}} IEEE Transactions on Information
  Theory}, 2004.

\bibitem{urbankecapacity}
TJ~Richardson and RL~Urbanke.
\newblock {The Capacity of Low-Density Parity-Check Codes Under Message-Passing
  Decoding}.
\newblock {\em IEEE Transactions on Information Theory}, 47(2):599--618, Feb
  2001.

\end{thebibliography}

\end{document}